\documentclass{acm_proc_article-sp}
\usepackage{url}
\usepackage{balance}
\usepackage{cite}
\usepackage{graphicx}
\usepackage{wrapfig}
\usepackage{hyphenat}
\usepackage{textcomp}
\usepackage{hyperref}

\begin{document}

%
\conferenceinfo{}{Bloomberg Data for Good Exchange 2017, NY, USA}

\title{Wildbook: Crowdsourcing, computer vision, and data science for conservation.}

%
%
%
%
%

\numberofauthors{10} 
%
\author{
%
%
\alignauthor
Tanya Y. Berger-Wolf\\
       \affaddr{University of Illinois at Chicago}\\
       \affaddr{Chicago, IL}\\
       \email{tanyabw@uic.edu}
\alignauthor
Daniel I. Rubenstein\\
       \affaddr{Princeton University}\\
       \affaddr{Princeton, NJ}\\
       \email{dir@princeton.edu}
\alignauthor Charles V. Stewart\\
       \affaddr{Rensselaer Polytechnic Inst.}\\
       \affaddr{Troy, NY}\\
       \email{stewart@cs.rpi.edu}
\and  
\alignauthor Jason A. Holmberg\\
       \affaddr{WildMe.org}\\
       \affaddr{Portland, OR}\\
       \email{jason@whaleshark.org}
\alignauthor Jason Parham\\
       \affaddr{Rensselaer Polytechnic Inst.}\\
       \affaddr{Troy, NY}\\
       \email{parhaj@rpi.edu}
\alignauthor Sreejith Menon\\
       \affaddr{Bloomberg LP}\\
       \affaddr{New York, NY}\\
       \email{smenon59@bloomberg.net}
}
\additionalauthors{
 Jonathan Crall (Rensselaer Polytechnic Inst.,  {\texttt{crallj@rpi.edu}}),\\
 Jon Van Oast (WildMe.org,  {\texttt{sito.org@gmail.com}}),\\
 Emre Kiciman (Microsoft Research,  {\texttt{emrek@microsoft.com}}),\\
 Lucas Joppa (Microsoft Research,  {\texttt{lujoppa@microsoft.com}}).
}

\maketitle

\section{Introduction}
\begin{wrapfigure}{l}{0.2\textwidth}
\vspace{-1cm}
\begin{center}
    \includegraphics[width=0.25\textwidth]{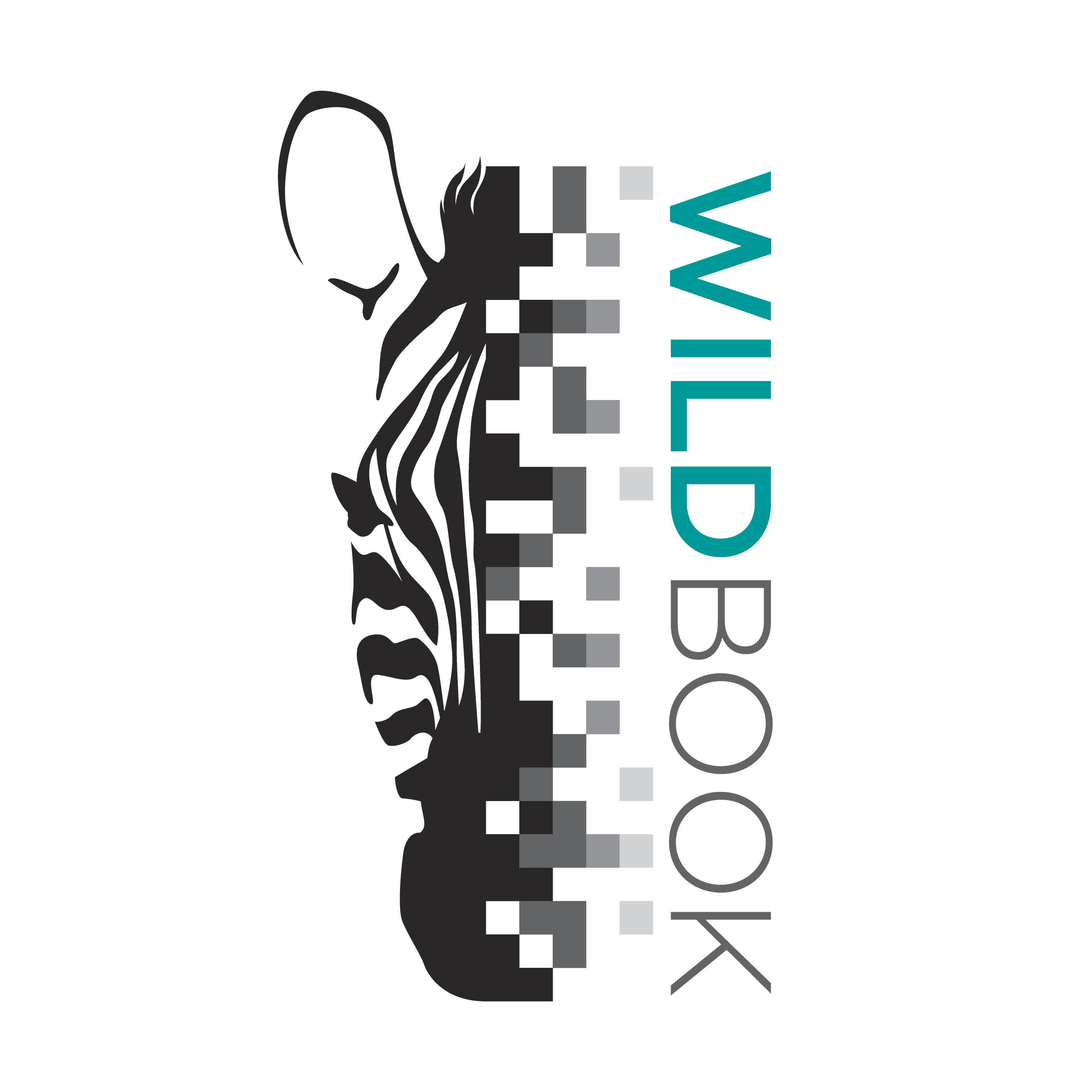}
\end{center}
\vspace{-1cm}
\end{wrapfigure}
How many African elephants are left in the world and how fast are they being lost to poaching? How far do whales travel? How many turtle hatchlings survive? Answers to these basic questions are critical to saving these and other endangered species and to the  conservation of the  biodiversity of our planet. However, this basic data is barely available for just a handful of species. The official body that tracks the conservation status of planet's species, International Union for Conservation of Nature (IUCN) Red List of Threatened Species\texttrademark\cite{IUCN}, currently has over 79,000 species~\cite{rodrigues2006value}. Yet, the Living Planet report, the most comprehensive effort to track the population dynamics of species around the world, includes just ~10,300 populations of just ~3,000 species~\cite{LivingPlanet}. That's not even 4\%! Scientists do not have the capacity to observe every species at the needed spatio-temporal scales and resolutions and there are not enough GPS collars and satellite tags to do so. Moreover, invasive tracking can be dangerous to the animals~\cite{OrcaInfectedTag}.

\begin{figure*}
	\centering
		\includegraphics[width=6.5in]{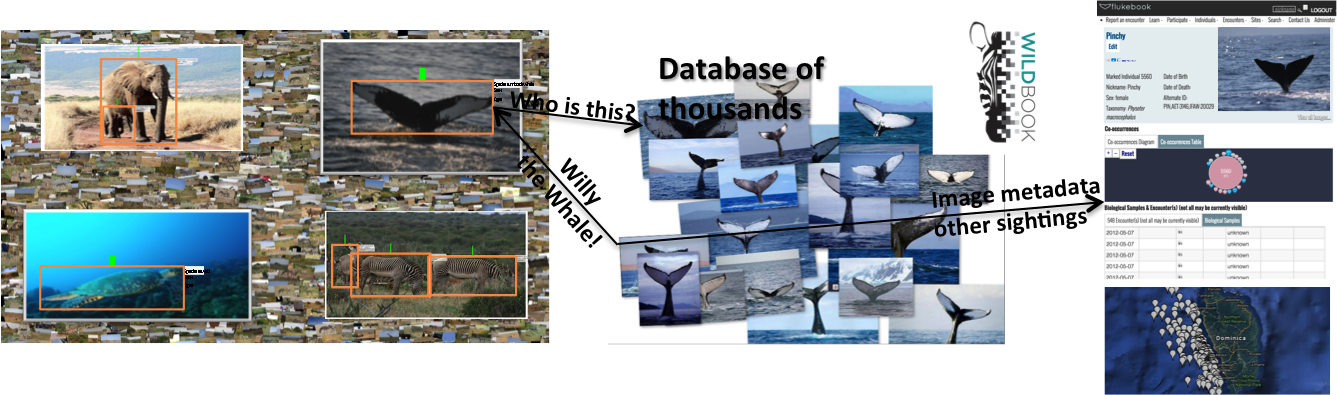}
	\caption{The Wildbook pipeline, starting with a collection of images, through species detection and individual animal identification, to the web-based data management layer and individual animal page.} 
	\label{fig:pipeline}
\end{figure*}

Images of animals and their environment, intentionally and opportunistically collected, are quickly becoming one of the  richest, most abundant, highest coverage and widely available source of data on wildlife. Coming from camera traps, cameras mounted on vehicles or UAVs (drones), photographs taken by tourists, citizen scientists, field assistants, scientists, and public photo streams,  many thousands of images may be collected per day from just one location.  Taking  advantage of this rich but big and messy source of data is only possible if we leverage computational approaches for every stage of the process, including image collection, information extraction, data modeling, and query processing.
%
We have developed algorithms and built a system, called {\bf Wildbook\texttrademark}\footnote{\url{http://Wildbook.org}}, based on the state of the art machine learning and data management approaches. Wildbook is an autonomous computational system that starts with an arbitrary heterogeneous collection of photographs of animals (Figure~\ref{fig:pipeline}). Wildbook can detect various species of animals in those images (using DCNN) and identify individual
animals of most striped, spotted, wrinkled or notched species~\cite{Crall2013}.  Wildbook can find matches within the
database or determine that the individual is new. Once an animal has been identified, it can be tracked in
other photographs.   It stores the information about who the animals
are and where and when they are there in a fully developed database, and provides query tools to that data for scientists researching population demographics, species distributions, individual interactions and movement patterns, as well as conservation managers and citizen scientists.~\cite{OReillyWildbookInterview}. 
 
\begin{figure*}
	\centering
		\includegraphics[width=6.5in]{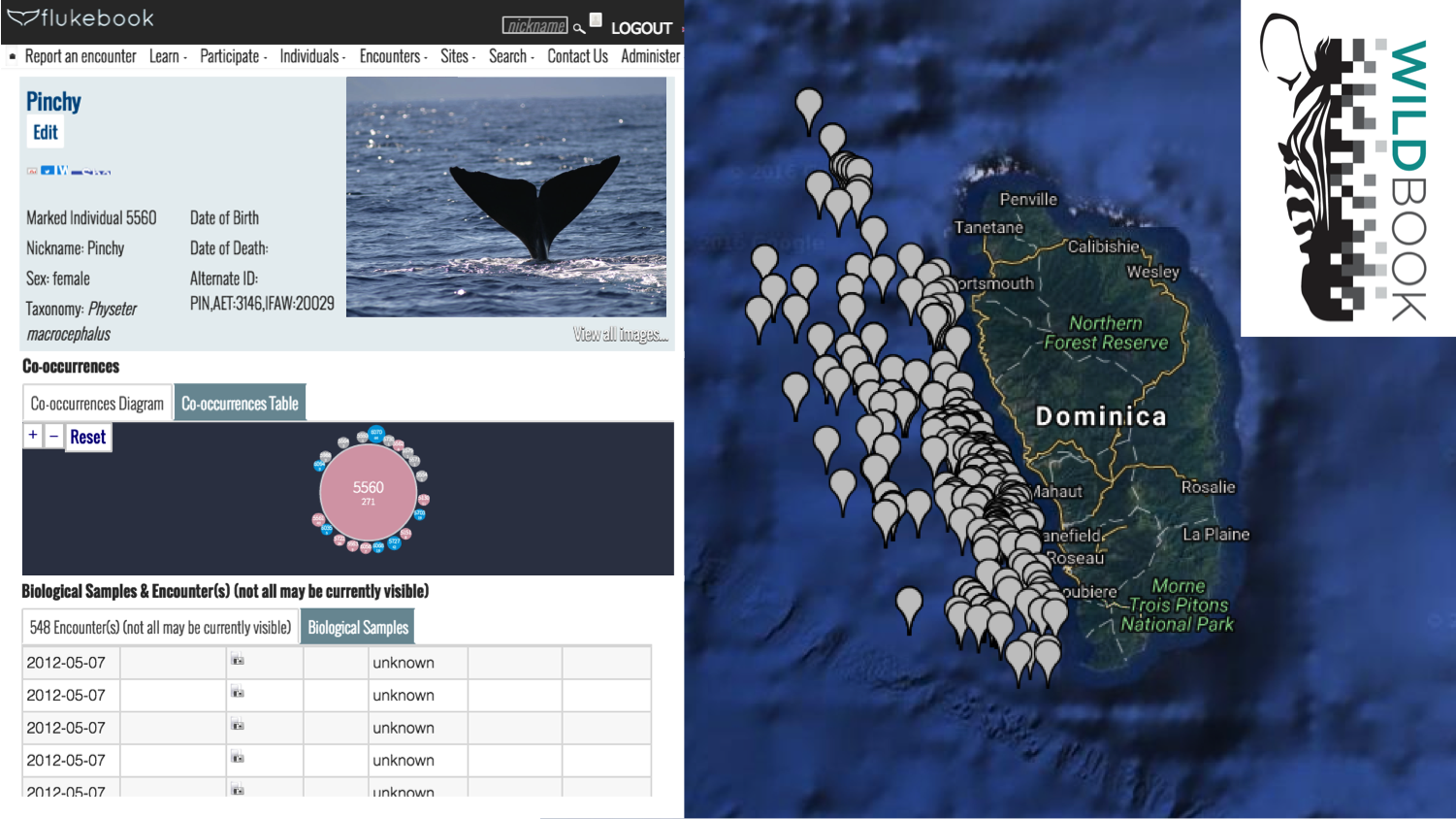}
	\caption{An example of a Flukebook (Wildbook for whales) page displaying information for an individual.} 
	\label{fig:flukebook}
\end{figure*}

Wildbook system allows to add biological data, as simple as sex and age, but also habitat and weather information, which allows to truly do population counts, birth/death dynamics, species range, social interactions or interactions with other species, including humans. An example of a Wildbook for whales, Flukebook, page is shown in Figure~\ref{fig:flukebook}.

Using Wildbook, it is possible to connect the information about sightings of animals ({\em who? where? when?}) derived from images to additional relevant data, providing the historic, current, and projected context of these sightings, thus enabling new science, conservation, and education, at unprecedented scales and resolution.  By layering additional data sets, covering everything from climate change and extreme weather to habitat ecology, agricultural development, urbanization, deforestation, the exotic animal trade, and the spread of disease, a much more detailed and useful picture of {\em what} is happening --- and {\em why} --- can be constructed within our architecture.

\section{Examples of Wildbook uses and impact}
Using our system, estimates of population sizes and movement patterns can be far more accurate, creating a better understanding of social structures and breeding of species, relationships between predators and prey, and  responses to environmental pressures, including land use by humans and long-term climate patterns.  Wildlife managers are better able to monitor the health of entire populations, discover dangerous trends, and reduce conflicts between humans and wildlife.  Access to information about individual animals, particularly visual information,  can also increase the public's understanding of the workings of science and its role in guiding conservation.  By contributing their photographs for scientific studies, visitors to parks and nature preserves in return learn the life histories of the individual animals they photograph and become connected to research projects and to the animals. We now present several examples of real impact a system like Wildbook can make in conservation policy, science, and public engagement.

\subsection{Evidence-based conservation policy: Lewa}
The first deployment of Wildbook was in January 2015 at Lewa Wildlife Conservancy\footnote{\url{http://www.lewa.org/}} in Kenya helping manage the endangered Grevy's zebra population.
The information from Wildbook for Grevy's showed that there are not enough babies surviving to adulthood mainly due to the lion predation. This lead to a change in the lion population management policy in Lewa helping save the endangered zebra.

\subsection{Crowdsourcing accurate conservation data: GZGC and GGR}
Knowing the number of individual animals within a population (a population census) is one of the most important statistics for research and conservation management in wildlife biology.  Moreover, a \textit{current} population census is often needed repeatedly over time in order to understand changes in a population's size, demographics, and distribution. This enables assessments of the effects of ongoing conservation management strategies.  Furthermore, the number of individuals in a population is seen as a fundamental basis for determining its conservation status.

Unfortunately, producing a population census is difficult to do at scale and across large geographical areas using traditional, manual methods.  One of the most popular and prevalent techniques for producing a population size estimate is mark-recapture~\cite{robson_sample_1964,pradel_utilization_1996} via a population count.  However, performing a mark-recapture study can be prohibitively demanding when the number of individuals in a population grows too large~\cite{seber_estimation_2002}, the population moves across large distances, or the species is difficult to capture due to evasiveness or habitat inaccessibility.  More importantly, however, a population \textit{count} is not as robust as a population \textit{census}; a count tracks sightings whereas a census tracks individuals.  A census is stronger because it can still produce a population size estimate implicitly but also unlocks more powerful ecological metrics that can track the long-term trends of individuals.  In recent years, technology has been used to help improve censusing efforts towards more accurate population size estimates~\cite{chase_continent-wide_2016,forrester_emammalcitizen_2014,simpson_zooniverse:_2014,swanson_snapshot_2015} and scale up\footnote{penguinwatch.org, mturk.com}.  However, these types of population counts are still typically custom, one-off efforts, with no uniform collection protocols or data analysis, and do not attempt to accurately track \textit{individuals} within a population across time.

To address the problems with collecting data and producing a scalable population census, we performed the following using Wildbook technology~\cite{Parham}: 
\begin{itemize}
    \item using citizen scientists~\cite{irwin_citizen_1995,cohn_citizen_2008} to rapidly collect a large number of photographs over a short time period (e.g.\ two days) and over an area that covers the expected population, and 
    \item using computer vision algorithms to process these photographs semi-automatically to identify all seen animals.
\end{itemize}

We showed that this proposed process can be leveraged at scale and across large geographical areas by analyzing the results of two completed censuses.  The first census is the Great Zebra and Giraffe Count (GZGC) held March 1-2, 2015 at the Nairobi National Park in Nairobi, Kenya to estimate the resident populations of Masai giraffes ({\em Giraffa camelopardalis tippelskirchi}) and plains zebras ({\em Equus quagga}).  The second is the Great Grevy's Rally (GGR) held January 30-31, 2016 in a region of central and northern Kenya covering the known migratory range of the endangered Grevy's zebra ({\em Equus grevyi}). 
While our method relies heavily on collecting a large number of photographs, it is designed to be simple enough for the average person to help collect them.  Any volunteers typically must only be familiar with a digital camera and be able to follow a small set of collection guidelines. 

The number of cars, volunteers, and the number of photographs taken for both rallies can be seen in Table~\ref{table:collection}.

\begin{table}
    \begin{center}
        \begin{tabular}{| l | l | l | l |}
            \hline
             & Cars & Cameras & Photographs\\ 
            \hline
            GZGC & 27 & 55 & 9,406\\ 
            \hline
            GGR & 121 & 162 & 40,810\\ 
            \hline
        \end{tabular}
    \end{center}
    \caption{The number of cars, participating cameras (citizen scientists), and photographs collected between the GZGC and the GGR.  The GGR had over 3-times as many citizen scientists who contributed 4-times the number of photographs for processing.}
    \label{table:collection}
\end{table}

Mark-recapture is a standard way of estimating the size of an animal population~\cite{chapman_fallow_1975,pradel_utilization_1996,robson_sample_1964}.  Typically, a portion of the population is captured at one point in time and the individuals are marked \textit{as a group}.  Later, another portion of the population is captured and the number of previously marked individuals is counted and recorded.  Since the number of marked individuals in the second sample should be proportional to the number of marked individuals in the entire population (assuming consistent sampling processes and controlled biases), the size of the entire population can be estimated.  

The population size estimate is calculated by dividing the total number of marked individuals during the first capture by the proportion of marked individuals counted in the second.  The formula for the simple Lincoln-Peterson estimator~\cite{pacala_population_1985} is:
\[
N_{\textrm{est}} = \frac{K*n}{k},
\]
where $N_{\textrm{est}}$ is the population size estimate, $n$ is the number of individuals captured and marked during the first capture, $K$ is the number of individuals captured during the second capture, and $k$ is the number of \textit{recaptured} individuals that were marked from the first capture.

\begin{table}
    \begin{center}
        \begin{tabular}{| l | l | l | l |}
            \hline
             & Annots. & Individuals & Estimate\\ 
            \hline
            GZGC Masai & 466 & 103 & 119 $\pm$ 4\\ 
            \hline
            GZGC Plains & 4,545 & 1,258 & 2,307 $\pm$ 366\\ 
            \hline
            GGR Grevy's & 16,866 & 1,942 & 2,250 $\pm$ 93 \\ 
            \hline
        \end{tabular}
    \end{center}
    \caption{The number of annotations, matched individuals, and the final mark-recapture population size estimates for the three species.  The Lincoln-Peterson estimate has a 95\% confidence range.  }
    \label{table:stats}
\end{table}

By giving the collected photographs to a computer vision pipeline, a semi-automated and more sophisticated census can be made.  The speed of processing large qualities of photographs allows for a more thorough analysis of the age-structure of a population, the distribution of males and females, and the movements of individuals and groups of animals, etc.  By tracking individuals, related to~\cite{jolly_explicit_1965,seber_note_1965}, our method is able to make more confident claims about statistics for the population.  The more individuals that are sighted \textit{and} resighted, the more robust the estimate and ecological analyses will be.\footnote{Portions of the results in this section were previously reported in two technical reports:~\cite{rubenstein_great_2015} for the GZGC and \cite{berger-wolf_great_2016} for the GGR.} The resulting estimates of the populations of Plains zebra and Maasai giraffe in Nairobi National Park and of the species census of the Grevy's zebra are the most accurate to date and the Grevy's zebra numbers are now used as the official numbers of the Grevy's zebra global population size by IUCN Red List~\cite{IUCNGrevys}.

\subsection{Crowdsourcing conservation data at scale: Whaleshark, Flukebook, online social media}
Today, Wildbooks for over a dozen species are available or are in the process of development. Wildbook for whales, Flukebook  (\url{http://flukebook.org/}), started with just over 800 individuals less than two years ago, is fully functional and helps track, protect, and study more than 11,600 marine mammals. Wildbook for whale sharks (\url{http://whaleshark.org/}) is the longest running animal sighting website which started with a couple of hundred individual animals 16 years ago as a volunteer effort and has more than 8,500 individuals with over 40,000 sightings today. IUCN Red List uses whaleshark.org for global population estimates~\cite{IUCNWhaleshark}. There are several projects with the WWF, World Wildlife Fund, with wildbooks for Saimaa Ringed seals (\url{http://norppagalleria.wwf.fi/}), lynx (\url{http://lynx.wildbook.org}), and sea turtles (the real IoT, Internet of Turtles: \url{http://iot.wildbook.org/}). As each Wildbook goes online, the number of identified individuals for each species grows over an order of magnitude within less than a year and Wildbook becomes the most reliable source of data for the species. Moreover, over the last year, we showed that social media can be a reliable source of information about animal populations~\cite{SWEEM} and Whaleshark.org uses public videos~\cite{WhalesharkYoutube}, while Flukebook is starting to use public images as a supplementary source of information.

\section{With Great Data comes Great Responsibility}
\subsection{Bias and accuracy}
Like all data, photographic samples of animal ecology are biased. These biases may lead to inaccurate population size or dynamic estimates. For example, even for such an iconic species as snow leopard, the last population estimate was in 2003 and ``many of the estimates are acknowledged to be rough and out of date"~\cite{IUCNSnowLeopard}. Yet, its conservation status can change depending on just a difference of a few individuals for some geographic locations. Human observers tend to overestimate population sizes since they may misidentify the same individual as different ones, while photo id provides evidence that those are indeed the same. Thus, image-based census can be used to support the more accurate population counts, which, in turn, will affect conservation status or policies for a species. That is indeed a big responsibility. 

To administer a correct population census, we must take these biases into account explicitly as different sources of photographs inherently come with different forms of bias. For example, stationary camera traps, cameras mounted on moving vehicles, and drones are each biased by their location, by the presence of animals at that location, by photographic quality, and by the camera settings (such as sensitivity of the motion sensor)~\cite{hodgson_unmanned_2013,hombal_multiscale_2010,foster_critique_2012,rowcliffe_clarifying_2013}.  These factors result in biased samples of species and spatial distributions, which recent studies are trying to overcome~\cite{ancrenaz_handbook_2012,maputla_calibrating_2013,xue_avicaching:_2016}. 

\begin{figure*}[h!]
    \centering
  \includegraphics[width=.8\textwidth]{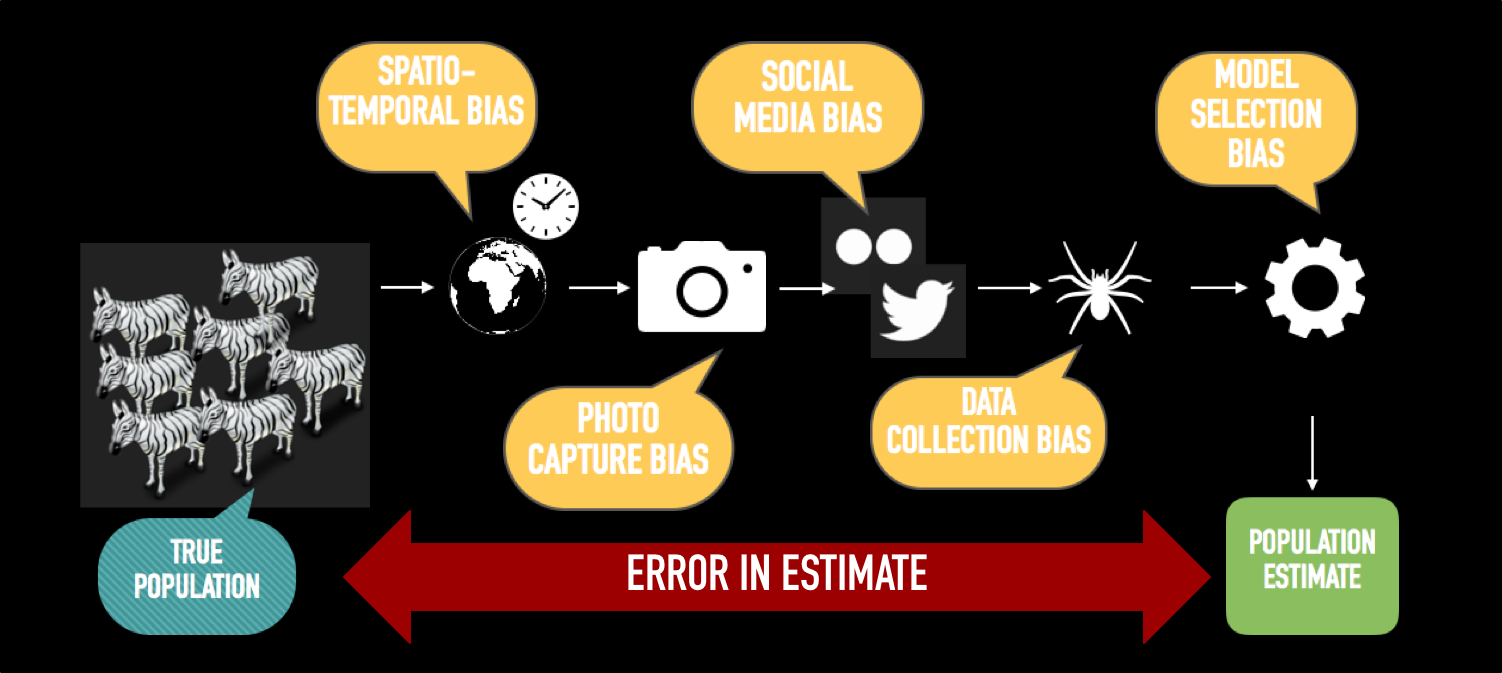}
  \caption{High level schematic representation of the problem of population estimation of wildlife species using images from social media, its challenges and biases in play.}
   \label{fig:bias}
\end{figure*}

Any human observer, including scientists and trained field assistants, is affected by observer bias~\cite{marsh_observer_2004,marsh_seeing_2007}.  Specifically, the harsh constraint of being at a single given location at a given time makes sampling arbitrary. Citizen scientists, as the foundation of the data collection, have additional variances in a wide range of training, expertise, goal alignment, sex, age, etc.~\cite{dickinson_citizen_2010}.  Nonetheless, recent ecological studies are starting to successfully take advantage of this source of data, explicitly testing and correcting for bias~\cite{van_strien_opportunistic_2013}; recent computational approaches address the question of if and how data from citizen scientists can be considered valid~\cite{wiggins_mechanisms_2011}, which can be leveraged with new studies in protocol design and validation. There are multiple biases that influences the final outcome of estimating population of a certain species from images that are obtained from social media. Some of the most prominent biases that influence the data we obtain from social media are outlined in Figure \ref{fig:bias}.  There are several layers of biases, accumulating in the resulting bias of estimating animal population properties from images. First, there are biases in the types of animals that people typically photograph in sufficient numbers in the first place. These may be charismatic or endangered species, or simply the ones easily observed. Second, there are biases in what images people take versus which ones they decide to share publicly on social media. These range from the Hawthorne Effect~\cite{Olteanu2016,bernstein20114chan,schoenebeck2013secret,shelton2015online} of changing behavior when knowing to be observed, to biases introduced by the demographics of the person sharing~~\cite{Olteanu2016,brenner2013demographics} and the choice of the social media platform~\cite{bruns2014twitter,gonzalez2014assessing,morstatter2014biased}. There are biases of our notions of beauty and aesthetics and cultural differences. Any mark-recapture model used to estimate the population size makes many assumptions and introduces its own biases. The fundamental question, however, is: {\em Do any of these actually affect the estimates of the population size and other parameters and if so, how?} Menon has begun to answer this question~\cite{Menon:MSThesis:2017} but a lot of work remains to be done. Moreover, combining these differently biased sources of data mutually constrains these biases and allows much more accurate statistical estimates than any one source of data would individually allow~\cite{bird_statistical_2014}.

\subsection{Security}
The technological sophistication of wildlife crime followed the increasing connectivity of remote conservation areas and availability of technology and data~\cite{PoachersDelight}. To mitigate some of the effects, nature conservancies warn tourists not to geotag their photos~\cite{NoGeotag} of endangered species but most forget or are anaware that they are doing it in the first place. The push for open research data is often in conflict with conservation, and new technologies such as animal recognition from photographs, add new challenges. Data security and privacy matters for endangered species, from snow leopards and elephants to pangolins and tortoises. Many of the issues are similar to the problems that humans face in an online globalized world, from insider threats to jurisdictional tangles, from multilevel policies to secondary-use hazards, from covert channels to geofencing, and from security economics to usability~\cite{PrivacyTigers}. Yet, in the context of wildlife conservation, there is no information security policy. It is urgent to address the issues of privacy and security for image-based wildlife data. To start, Wildbook has worked with the Center for Trustworthy Scientific Cyberinfrastructure to design a secure system to prevent data leakage and poachers' access yet more work remains to be done.

\section{Conclusions and challenges}

We have designed, implemented and deployed a prototype, and are continuing to develop Wildbook, an image-based ecological information system. Using image analysis algorithms and state-of-the-art information management infrastructure, Wildbook adds images, opportunistically and intentionally crowdsourced and scientifically collected, to the source of data about animals and provides the analytical tools to gain scientific and conservation insight from those data. As the new type of data, the images are not only augmenting the scale and resolution of existing scientific and conservation inference, but allow {\em new types} of questions that lead to new scientific understanding of why animals do what they do, as well as a change in the conservation policy. Moreover, we have already demonstrated that Wildbook provides a platform and a tool for a much more personal and committed public engagement in science and conservation than has been available to date. By enabling events such as the Great Zebra and Giraffe Count and Great Grevy's Rally, Wildbook both presents an instant, easily available, no-training-needed route for general public contribution to science and conservation, as well as creating a personal bond with animals and nature by providing an instant individual animal identity. Moreover, it provides data for evidence-based conservation policy at large scale and high resolution over time, space, and individual animals.

However, to achieve these new insights and engagement, many challenges need to be overcome. In addition to the many computational, scientific, and societal challenges, there are two directly related to data. First, Wildbook requires a new infrastructure that can function, synchronize, and coordinate across many platforms, from the mobile phones and GPS cameras of the citizen scientists, the bandwidth and electricity-starved research stations and conservation outposts in remote uninstrumented locations, to the cloud infrastructure  containing information about entire species and regions, as well as the algorithms necessary for its analysis. The data-related aspects of this infrastructure challenges are about information aggregation, integration, synchronization, semantic complexity, and access control.  The second big data-related challenge that the new data sources and the enabled use of those data present is the unknown data biases that challenge traditional computational methods and analytical tools. From the simplest population size and species range estimates, the traditional methods rely on a uniform random sampling regime. While it is not clear that the assumption is true for any of the data collection methods, it is most definitely does not hold for the data coming from citizen scientists' and tourists' photographs. Our preliminary analysis shows that there are wide variations in the number and rate of images taken and complex patterns of camera and species fatigue that arise from demographic, cultural, and event-specific factors. Accounting for or leveraging those in designing the new generation of analytical tools is a challenge and a goal of Wildbook and it is critical for reliable conservation policy decisions.

Finally, one potential use of Wildbook that presents very different data-related challenges is as a tool in wildlife crime prevention. Wildbook ability to track individuals through photographs during their life, as well as identifying these individuals by a reasonably sized part of the body later, given that part has been previously photographed, allows photographic evidence to be used both in forensics and as a deterrence in poaching, killing, and illegal trafficking of animals. Wildlife crime is threatening to wipe out many charismatic species from the planet: rhino population (across species) is down 90\% from its height~\cite{RhinoNumbers} and 100,000 elephants were killed over the last 3 years in Africa for ivory~\cite{Wittemyer09092014}. Many charismatic fauna species, such as leopards, elephants, tigers, zebras, snow leopards, turtles, and tortoises, have individual-level uniquely identifiable body patterns, essentially `body-prints'. With an Wildbook app, a law enforcement official would be able to take a picture of an animal crime victim (alive or not) and be able to find a match in the reference database if one exists. Thus, the identity, geographic origin and life history of the animal will be instantly available. There are two types of primary users. The first are conservation and wildlife managers responsible for overseeing a particular endangered species of identifiable fauna. They collect photographs and submit information to Wildbook to build up the databases. The second type of users are law enforcement officials who would take pictures of animals or their hides or carcasses, submit these to Wildbook to obtain, if known, the identity and origin of the specimen. 

The use of Wildbook or a similar information system  for the purposes of conservation of highly endangered species or wildlife crime prevention presents a unique problem in data security and privacy. Ironically, every new technology is a ``double-edged sword" for wildlife and is often used by the criminals to aid in illegal wildlife trade~\cite{PoachersDelight}. Thus, the location, current or predicted, of valuable or highly endangered species or individuals must be protected. Privacy and security protocols must be developed to protect animal data. The benefits of opening data for science, conservation, and public engagement must be weighed against the threat of species extinction.

\section*{Acknowledgements}
This work was supported in part by a gift from Microsoft Research, funding from the Department of Computer Science at the University of Illinois at Chicago, and NSF grants CNS-1453555 EAGER: Prototype of an Image-Based Ecological Information System (IBEIS) and EF-1550853 EAGER-NEON: Image-Based Ecological Information System (IBEIS) for Animal Sighting Data for NEON. We would also like to acknowledge in-kind support, volunteer time, and collaborations with \href{http://www.mpala.org/}{Mpala Research Centre} (Kenya), \href{http://www.lewa.org/}{Lewa Wildlife Conservancy} (Kenya), \href{http://WildlifeDriect.org}{WildlifeDriect.org}, \href{http://www.grevyszebratrust.org/}{Grevy's Zebra Trust}, and \href{http://www.kws.go.ke/}{Kenya Wildlife Service}.

\balance
\bibliographystyle{abbrv}
\bibliography{references,referencesSWEEM}
\end{document}